\documentclass[11pt,a4paper]{article}

\usepackage[left=2cm,top=2cm,right=2cm,bottom=2cm,nohead]{geometry}

\usepackage{a4wide}
\usepackage{graphicx}
\usepackage[T1]{fontenc}
\usepackage{mathptmx}
\usepackage{MnSymbol}
\usepackage{pifont}
\usepackage{amsmath,amsthm}
\usepackage{cite}
\usepackage{xcolor}
\usepackage{soul}
\usepackage[font=small]{caption}

\usepackage{hyperref}

\graphicspath{{./}{./figures/}{./figures/model_selection/}}

\newcommand{\findr}{\textit{Findr}}

\newcommand{\ced}[1]{{#1}}

\newcommand{\cedst}[1]{}

\title{Comparison between instrumental variable and mediation-based methods for reconstructing causal gene networks in yeast}

\author{Adriaan-Alexander Ludl and Tom Michoel$^*$}

\begin{document}

\maketitle

Computational Biology Unit, Department of Informatics, University of Bergen, PO Box 7803, 5020 Bergen, Norway

$^*$ Corresponding author, email: \texttt{tom.michoel@uib.no}

\begin{center}
  \textbf{Abstract}
\end{center}

Causal gene networks model the flow of information within a cell. Reconstructing causal networks from omics data is challenging because correlation does not imply causation. When genomics and transcriptomics data from a segregating population are combined, genomic variants can be used to orient the direction of causality between gene expression traits. Instrumental variable methods use a local expression quantitative trait locus (eQTL) as a randomized instrument for a gene's expression level, and assign target genes based on distal eQTL associations. Mediation-based methods additionally require that distal eQTL associations are mediated by the source gene. A detailed comparison between these methods has not yet been conducted, due to the lack of a standardized implementation of different methods, the limited sample size of most multi-omics datasets, and the absence of ground-truth networks for most organisms. Here we used Findr, a software package providing uniform implementations of instrumental variable, mediation, and coexpression-based methods, a recent dataset of 1,012 segregants from a cross between two budding yeast strains, and the YEASTRACT database of known transcriptional interactions to compare causal gene network inference methods. We found that causal inference methods result in a significant overlap with the ground-truth, whereas coexpression did not perform better than random. A subsampling analysis revealed that the performance of mediation \ced{saturates} at large sample sizes, due to a loss of sensitivity when residual correlations become significant. Instrumental variable methods on the other hand contain false positive predictions, due to genomic linkage between eQTL instruments. Instrumental variable and mediation-based methods also have complementary roles for identifying causal genes underlying transcriptional hotspots. Instrumental variable methods correctly predicted \textit{STB5} targets for a hotspot centred on the transcription factor \textit{STB5}, whereas mediation failed due to Stb5p auto-regulating its own expression. Mediation suggests a new candidate gene, \textit{DNM1}, for a hotspot on Chr XII, whereas instrumental variable methods could not distinguish between multiple genes located within the hotspot. In conclusion, causal inference from genomics and transcriptomics data is a  powerful approach for reconstructing causal gene networks, which could be further improved by the development of methods \ced{to control for residual correlations in mediation analyses and}  genomic linkage and pleiotropic effects from transcriptional hotspots \ced{in instrumental variable analyses}.

\newpage

\section{Introduction}
\label{sec:introduction}

Causal gene networks model the flow of information from genotype to phenotype within a cell or whole organism \cite{jansen2001genetical,rockman2008reverse,schadt2009,boyle2017expanded}. Reconstructing causal networks from omics data is challenging because correlation does not imply causation. However, when genomics and transcriptomics data from a large number of individuals in a segregating population are combined, genomic variants can be used to orient the direction of causality between gene expression traits. This is based on the fact that alleles are randomly segregated during meiosis and genotypes remain fixed during an individual's lifetime, such that genomic variants act as causal anchors from which all arrows are directed outward \cite{jansen2001genetical,rockman2008reverse,li2010critical}. Moreover, local and distal expression quantitative trait locus (eQTL) associations have biologically distinct interpretations, because genomic variation at regulatory DNA elements leads to altered transcription of nearby genes by \textit{cis}-acting, epigenetic mechanisms, whereas distal associations must be intermediated by \textit{trans}-acting factors \cite{albert2015role, pai2015genetic}

These principles are combined in different ways in two classes of causal inference methods that use genomic variants as causal anchors: \emph{instrumental variable} \ced{(known as \emph{Mendelian randomization} in genetic epidemiology)} or \emph{mediation}-based  \cite{hemani2017orienting}. \emph{Mediation} infers the direction of causality between two traits that are statistically associated to the same genomic variant by testing whether the association between the variant and one of the traits is mediated by the other trait, in which case there must be a causal relation from the mediating trait to the other one \cite{schadt2005integrative,chen2007harnessing}. Mediation does not require that one of the traits has a ``preferential'' relation to the genomic variant (as in \textit{cis} or \textit{trans}).  However, mediation fails in the presence of high measurement noise or hidden confounders, such as common upstream factors coregulating both traits, where it  rejects true interactions (i.e.\ reports false negatives) \cite{wang2017}.

\emph{Instrumental variable} \ced{or \emph{Mendelian randomization}} methods assume that the genomic variant acts as a randomized ``instrument'' for one of the traits, similar to the random assignment of individuals to treatment groups in randomized controlled trials, such that a statistical association between the variant and the second trait is evidence for a causal relation from the first to the second trait. The random group assignment, in genetics the random segregation of alleles, ensures that causal effects can be detected even in the presence of confounding. However, instrumental variable methods fail if there are pathways from the variant to the second trait other than through the first trait (pleiotropic effects) \cite{davey2003mendelian,lawlor2008mendelian,davey2014mendelian}.

A detailed comparison between these two approaches requires a standardized implementation where pre-processing (e.g. data normalization) and post-processing (e.g. multiple testing correction) are handled uniformly. Previously, we developed \findr, a computationally efficient software package implementing six likelihood ratio tests that can be combined in multiple ways to reconstruct instrumental variable as well as mediation-based causal gene networks \cite{wang2017}. \findr\ expresses the result of each test as a posterior probability (one minus the local false discovery rate), allowing tests to be combined by the usual rules of probability theory \cite{chen2007harnessing}. \ced{This results in causal network inference methods that are representative for the broader field. For instance, the implementation of the mediation-based method in \findr\ is identical to the method of Chen \emph{et al.}\cite{chen2007harnessing}, which had its roots in the ``likelihood-based causal model selection'' (LCMS) procedure of Schadt \emph{et al.}\cite{schadt2005integrative}. The Causal Inference Test (CIT) software \cite{millstein2009disentangling,millstein2016cit} is another implementation of an LCMS-based mediation method, which combines statistical tests using omnibus $p$-values and FDR estimates. We found previously that it results in similar inferences as the mediation-based method implemented in \findr \cite{wang2017}. Instrumental variable methods on the other hand are based on genetic associations, for which \findr\ uses categorical regression of gene expression profiles on genotype values, similar to for instance the ANOVA option in Matrix-eQTL \cite{shabalin2012matrix}.}

Using simulated data from the DREAM5 Systems Genetics challenge \cite{pinna2011simulating,marbach2012wisdom}, we found previously that instrumental variable methods generally outperformed mediation-based methods in terms of area under the precision-recall curve, and that the performance of mediation-based methods \emph{decreased} with increasing sample size, due to increased statistical significance of confounding effects \cite{wang2017}. However, at that time, no real-world dataset with sufficient sample size as well as an accurate ground-truth network of causal interactions was available to test these predictions in a real biological system.

Fortuitously, a dataset has now become available of genomic variation and gene expression data in more than 1,000 segregants from a cross between two strains of budding yeast, a popular eukaryotic model organism \cite{albert2018genetics}. By learning networks from these data, and comparing against the wealth of transcriptional regulatory interactions and other functional validation data available for budding yeast \cite{monteiro2020yeastract}, a thorough benchmarking of methods for reconstructing causal gene networks has become possible.

\section{Methods}
\label{sec:methods}

\subsection{Selecting strongest \emph{cis}-eQTLs}
\label{sec:eqtls}

Using the data on expression quantitative trait loci (eQTLs)
from \cite{albert2018genetics}, we selected the strongest \textit{cis} acting eQTLs for 2884 genes.
The eQTLs were ranked in descending order according to the absolute value of the correlation  coefficient  between scaled expression levels and marker genotype ($r$, obtained from \cite[Source data 4]{albert2018genetics}),
and for each gene the highest ranked eQTL was retained.
Among the selected eQTLs 2044 occured once, 337  eQTLs were strongest for two genes,
 44 were strongest for three genes, 
 6 were strongest for four genes, 
 2 were strongest for five genes.

\subsection{Network inference methods}
\label{sec:infer}

We used the inference methods implemented in \findr\ \cite{wang2017}.
The source code is available at \url{https://github.com/lingfeiwang/findr}.
The test $P_0$ only uses gene expression data.
For the other tests ($P$, $P_2$, $P_3$, $P_5$),
 we used the genotype and gene expression data from \cite{albert2018genetics}
 (see  section \ref{sec:data} below for details)
  with \emph{cis}-eQTLs as causal anchors for the
inference tests.
Composite tests are obtained by element-wise multiplication of the matrices containing the results of individual tests.

\subsection{Performance measures}
\label{sec:performance}

The Precision-Recall curves and area under the curve (AUPR) for interactions predicted by a given test were computed using the scikit-learn package \cite{skl2011} and three ground-truth matrices (see Data section \ref{sec:data}). Recall is equivalent to the true positive rate (TPR), i.e.\ the number of true positive predictions as a fraction of all known positive interactions in the network. Precision or positive predictive value is $1-\mathrm{FDR}$ where FDR is the global false discovery rate.
 
AUPR-ratio or fold-change is the AUPR divided by the theoretical value for random predictions on a given ground truth.
The latter is obtained as the precision for random predictions given by
    $prec_\mathrm{random} = N_E /( N_R * N_T)$
    where $N_R$ is the number of regulating genes, 
     $N_T$ is the number of target genes, 
     $N_E$ is the number of edges, i.e.\ the number of ones in the ground-truth adjacency matrix.

\subsection{Subsampling}
\label{sec:subsamples}

We performed subsampling on the segregants to evaluate the change in performance of our inference methods on various sample sizes. Four subsamples of randomly selected segregants were drawn for the following sizes: 10, 100, 200, 400, 600, 800 and 1,000. The inference methods were run on each sample.
We report the average AUPR and its statistical standard deviation over the four subsamples in Fig. \ref{fig:subsampling}.

\subsection{Genotypes covariance and target counts}
\label{sec:covar}

We computed the covariance matrix of the genotypes at the retained eQTLs for all 1,012 segregants (Fig. \ref{fig:hotspots}).
The rows and columns of the matrix were reordered according to the genome position of the eQTL, the ordering algorithm is described below in \ref{sec:data}.

\findr\ posterior probability matrices were thresholded to obtain discrete networks with an expected FDR target value as described previously \cite{wang2017,chen2007harnessing}: because
for each interaction the local false discovery rate  is given by $\mathrm{fdr} = 1-p$, where $p$ is the posterior probability value obtained by the test, the expected FDR of a network consisting of all interactions with $p\geq p_\mathrm{th}$ is the average of the local fdr of the retained interactions. We determined $p_\mathrm{th}$ as the threshold that gave the greatest expected FDR below the target value (5 or 10 \%).
We counted the number of targets for each source gene whose $p>p_\mathrm{th}$.

\subsection{Software and Data availability}
\label{sec:data}

\ced{The inferred regulatory relationships for the thresholds reported in Tab. \ref{tab:network_properties} for the causal tests ($P_2P_3$, $P_2$, $P_2P_5$, $P$) and scripts to reproduce the analysis are provided in the repository \url{https://github.com/michoel-lab/FindrCausalNetworkInferenceOnYeast}.
Running all Findr inference tests on the data from \cite{albert2018genetics} takes about 10 to 15 seconds on a
typical desktop computer.}

\subsection{Data}
\label{sec:data}

We used gene expression data for 5,720 genes and genotypes
for a panel of 1,012 segregants from crosses of one laboratory strain (BY) and a wine strain (RM)
 from \cite{albert2018genetics}.
\ced{Batch and optical density (OD) effects, as given by the covariates provided in} \cite{albert2018genetics},
     were removed from the expression data using categorical regression,
     as implemented in the statsmodels python package \cite{seabold2010statsmodels}.
The paper also provides data on expression quantitative trait loci (eQTLs)
that was used to select the strongest \emph{cis}-eQTLs, as well as a file with annotations to the 102 hotspots that they identified.

For validation we used networks of known transcriptional regulatory
interactions in yeast (S. cerevisiae)  from \textsc{Yeastract}  
\cite{monteiro2020yeastract}.
Regulation matrices were obtained from \url{http://www.yeastract.com/formregmatrix.php}.
We retrieved the full ground-truth matrices containing all reported interactions of the following types from the \textsc{Yeastract} website:
     \emph{DNA binding evidence} was used as the ``Binding'',
     \emph{expression evidence}  including 	TFs acting as activators and those acting as inhibitors  was used as the ``Expression'',
     \emph{DNA binding and expression evidence} was used as the ``Binding \& Expression''.
     Self regulation was removed from all ground truths.
     The numbers of regulators, targets and interactions for these three ground-truth networks are shown in Tab. \ref{tab:ground_truth}.

\begin{table}[th]
\begin{center}
\begin{tabular}{lrrrr}
\\\hline  Ground-Truth Network  &  $N_R$    & $N_T$  & $N_E$    & $N_{sE}$
\\\hline	Binding	            &	90	    & 5,151	 & 19,099   & 28
\\	        Binding\&Expression	&	80	    & 3,394	 & 5,680    & 24
\\	        Expression	        &	113	    & 5,369	 & 92,646   & 77
\\\hline
\end{tabular}
\end{center}
\caption{\textbf{Properties of the \textsc{Yeastract} ground-truth networks.}
    $N_R$ is the number of regulating genes, 
     $N_T$ is the number of target genes, 
     $N_E$ is the number of edges excluding self-edges, 
     $N_{sE}$ is the number of self-edges.
    Data was retrieved from \textsc{Yeastract} \cite{monteiro2020yeastract}.
    }
\label{tab:ground_truth}
\end{table}

Annotations of the yeast genome were used to map gene names to their identifiers 
and order them according to the position of their causal anchor (eQTL) along the full genome, first by chromosome and then by position along the chromosome.
The sorting algorithm places mitochondrial genes first (when present) and orders the chromosomes according to the numerical value of the roman numerals.
We used the gff3 file ( \path{Saccharomyces_cerevisiae.R64-1-1.83.gff3.gz} ) from the Ensembl database (release 83, December 2015),
\cite{ensembl2020},
which is the version used by \cite{albert2018genetics}.
The file is accessible at \url{ftp://ftp.ensembl.org/pub/release-83/gff3/saccharomyces_cerevisiae/}.

\section{Results}
\label{sec:results}

\subsection{Findr reconstructs instrumental variable and mediation-based causal gene networks in yeast}
\label{sec:findr-reconstr}

\begin{figure}[ht!]
  \centering
  \includegraphics[width=\linewidth]{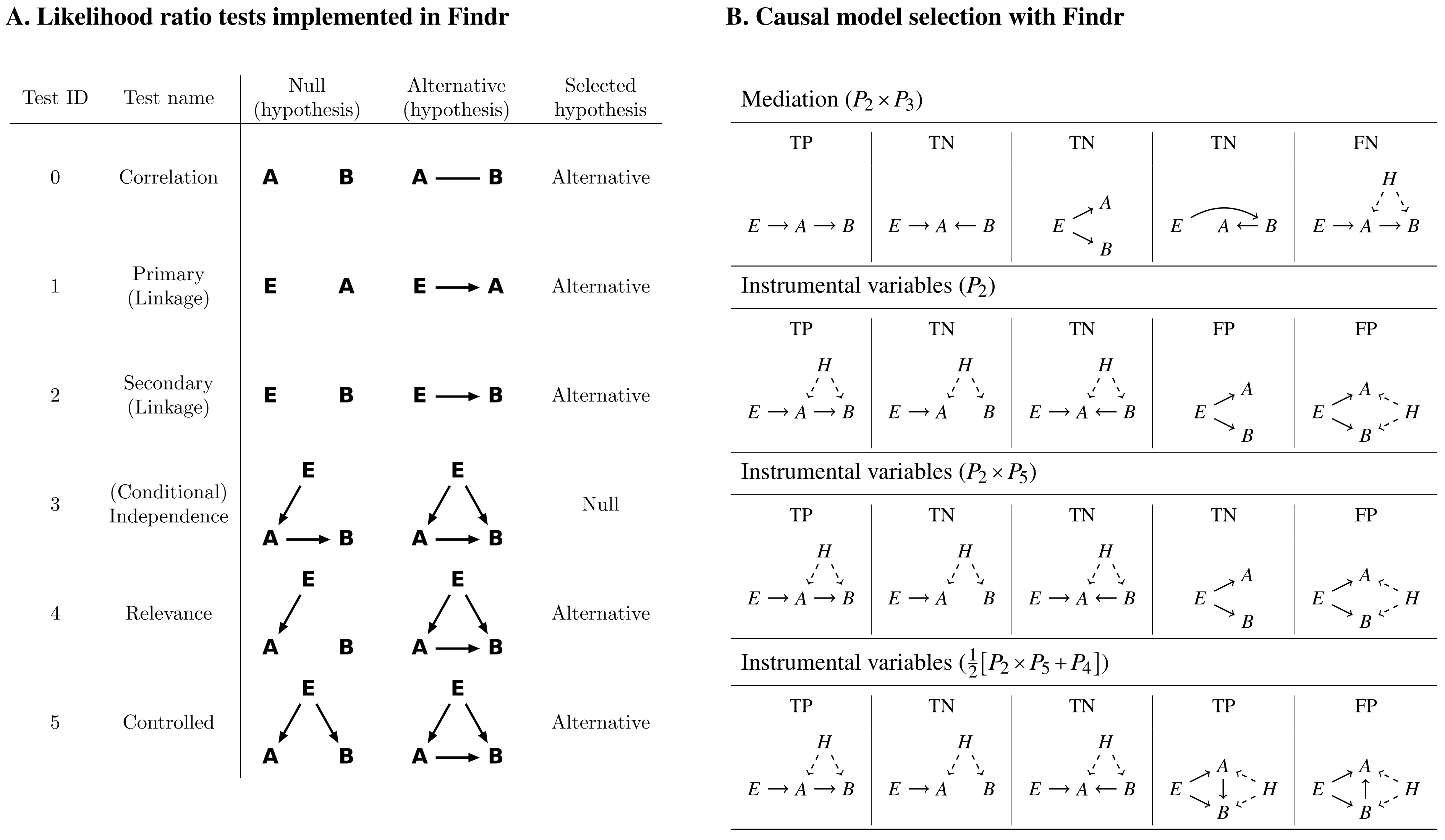}
  \caption{\textbf{A. Likelihood ratio (LLR) tests implemented in Findr.} $E$ is a causal anchor of gene $A$. Arrows in a hypothesis indicate directed regulatory relations. 
  Genes $A$ and $B$ each follow a normal distribution, whose mean depends additively on its regulator(s), as determined in the corresponding hypothesis. The dependency is categorical on genotypes and linear on gene expression levels. The undirected line represents a multi-variate normal distribution between the relevant variables. In each test, either the null or the alternative hypothesis is selected, as shown. Figure \copyright 2017 Wang, Michoel,  reproduced by permission from \cite{wang2017} under Creative Commons Attribution License. \textbf{B. Causal model selection with Findr.} By combining the posterior probabilities $P_i$ of the selected hypothesis for test $i$ being true, Findr determines whether coexpressed genes $A$ and $B$ are connected by a causal $A\to B$ relation. \textbf{(Row 1)} In the absence of hidden confounders ($H$), mediation-based causal inference, combining Findr tests 2 and 3, correctly identifies true positive (TP; correlation due to causal $A\to B$ relation) and true negative (TN; correlation without causal $A\to B$ relation) models. However, it reports a false negative (FN) if the causal relation is affected by a hidden confounder. \textbf{(Row 2)} If the causal anchor is ``exclusive'' to gene $A$, then the instrumental variable method based on Findr test 2 correctly identifies TP and TN models, even in the presence of hidden confounding. However, it reports a false positive (FP) if the association between $E$ and $B$ is due to other paths than through $A$ (pleiotropy). \textbf{(Row 3)} An  instrumental variable method that combines Findr tests 2 and 5 correctly identifies a true negative if the correlation between $A$ and $B$ is entirely due to a pleitotropic effect of $E$, but will still report a false positive if there is an additional effect from a hidden confounder. \textbf{(Row 4)} An instrumental variable method based on the compound hypothesis that test 4 is true, or test 2 and test 5 are true, reports a TP for causal relations where $E\to A\to B$ is not the only path from $E$ to $B$, with or without confounding, but will report a FP if the true causal relation is $B\to A$ (or absent).}
  \label{fig:tests-models}
\end{figure}

We used the software \findr\ \cite{wang2017} to reconstruct causal and non-causal gene networks in yeast from a dataset of genomic variation and expression data for 5,720 genes in  1,012 segregants from a cross between two strains of budding yeast \cite{albert2018genetics}. 2,884 genes had an associated genomic causal anchor, here defined as the variant most strongly associated to the gene and present in the list of genome-wide significant eQTLs
whose confidence interval (of variable size) overlaps with an interval covering the gene, 1,000 bp upstream and 200 bp downstream of the gene position \cite{albert2018genetics}. \findr\ implements six likelihood ratio (LLR) tests between triplets $(E,A,B)$, where $A$ and $B$ are genes, and $E$ is the causal anchor for $A$. For each test $i$, \findr\ outputs the posterior probability $P_i$ of the selected hypothesis being true (Fig.~\ref{fig:tests-models}A). These posterior probabilities can then be combined to obtain the posterior probabilities of various compound hypotheses being true. Here we considered four causal tests and one non-causal test to reconstruct directed gene networks:
\begin{itemize}
\item \emph{Mediation.} Mediation-based approaches infer a causal interaction $A\to B$ if gene $B$ is statistically associated to the causal anchor $E$, and the association is abolished after conditioning on gene $A$ \cite{schadt2005integrative,chen2007harnessing,millstein2016cit}. In \findr\ this is accomplished by the compound hypothesis that test 2 and 3 are both true, i.e. by the posterior probability $P_2\times P_3$. Mediation can distinguish true positive (TP) from true negative (TN) causal interactions in the absence of hidden confounders, but will report a false negative (FN) if a real causal interaction is confounded by a hidden factor (Fig.~\ref{fig:tests-models}B, row 1), due to a collider effect \cite{chen2007harnessing,wang2017}.
\item \emph{Instrumental variables without pleiotropy.} Instrumental variable approaches assume that the causal anchor $E$ acts as a randomized instrument for gene $A$, and, in their simplest form, infer a causal interaction $A\to B$ if gene $B$ is statistically associated to the causal anchor $E$, i.e.\ by the posterior probability $P_2$ that test 2 is true. Instrumental variables can distinguish true positive from true negative causal interactions even in the presence of hidden confounders, but will report a false positive (FP) if there are other pathways than through $A$ that cause a statistical association between $E$ and $B$ (pleiotropy) (Fig.~\ref{fig:tests-models}B, row 2).
\item \emph{Instrumental variables with perfect pleiotropy.} To address the problem of pleiotropy, we can additionally require that genes $A$ and $B$ are not independent after conditioning on $E$, accomplished by the compound hypothesis that test 2 and 5 are both true, i.e. by the posterior probability $P_2\times P_5$. This correctly identifies a true negative if $E$ explains all of the correlation between $A$ and $B$, but will still result in a false positive if there is a hidden confounder (Fig.~\ref{fig:tests-models}B, row 3).
\item \emph{Instrumental variables with partial pleiotropy.} To overcome the problem of FP predictions in the ``confounded pleiotropy'' situation, we introduced test 4 in \findr, which tests whether gene $B$ is not independent of $E$ and $A$ simultaneously, and found empirically that the combination $P=\frac12(P_2P_5 + P_4)$ performs better than $P_2\times P_5$ alone \cite{wang2017}. In particular, it identifies a TP for causal $A\to B$ relations even in the presence of alternative $E\to B$ paths and hidden confounding, at the expense of FP predictions when the relation is reversed or absent (Fig.~\ref{fig:tests-models}B, row 4).
\item \emph{Coexpression.} As a basic reference, we reconstructed a gene network based on coexpression alone, using \findr\ test 0. Note that the posterior probability $P_0$ is not symmetric ($P_0(A\to B)\neq P_0(B\to A)$), because it is estimated from the observed distribution of LLR test statistics for each $A$ separately \cite{wang2017}.
\end{itemize}

\begin{figure}[ht]
\begin{center}
\includegraphics[width=.95\textwidth]{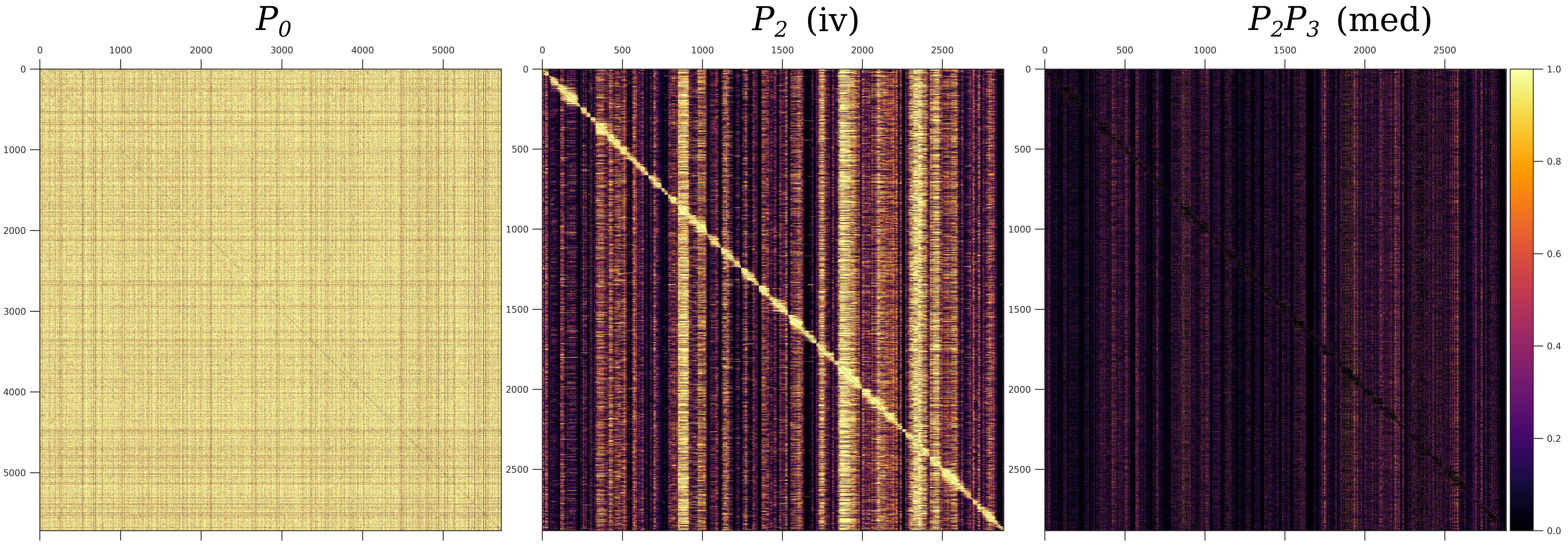}
\end{center}
\caption{\textbf{Matrices of predicted gene interactions.}
    These square matrices represent the interactions between 2884 genes with causal anchors (eQTLs),
        posterior probability values are color coded.
        Vertical bands correspond to hotspots.
    \textbf{Left}: The correlation based test $P_0$.
    \textbf{Center}: The instrumental variable test $P_2$.
    \textbf{Right}: The mediation test $P_2P_3$.
    The genes are ordered according to the position of their causal anchor in the full yeast genome. See Sup.\ Fig.~\ref{fig:matrices_sup} for the instrumental variable tests $P_2P_5$ and $P$.
    }
\label{fig:matrices}
\end{figure}

To illustrate the differences between coexpression, instrumental variable, and mediation-based gene networks, we considered the sub-networks inferred between the 2,884 genes that had a causal anchor (i.e. the sub-network where the probability of an edge can be estimated for \emph{both} edge directions). As expected, the coexpression network ($P_0$) is largely symmetric (Fig.~\ref{fig:matrices}, left), whereas the causal instrumental variable ($P_2$, Fig.~\ref{fig:matrices}, center) and mediation-based ($P_2 P_3$, Fig.~\ref{fig:matrices}, right) networks show a clear asymmetric structure with some genes having a very large number of high-confidence targets. These genes correspond to transcriptional hotspots, regions of the genome with a large, genome-wide effect on gene expression \cite{albert2018genetics}. The overall structure of the causal networks appears consistent with the general considerations above. The overall signal (strength of posterior probabilities) is weaker in the mediation-based network, consistent with an increased false negative rate (Fig.~\ref{fig:matrices}, right). On the other hand, the instrumental variable network appears to have a genomic structure, where nearby genes are mutually connected and have a similar target profile (Fig.~\ref{fig:matrices}, middle). This \cedst{is consistent with pleiotropic effects where} \ced{could be due to} genomic linkage between causal anchors\ced{: if two genes $A$ and $A'$ share the same or highly correlated instruments $E$ and $E'$, then their predicted target sets would also be very similar, and probably include a large proportion of} \cedst{would lead to} false positive predictions \ced{for either gene}.

\subsection{Causal gene networks overlap significantly with known transcriptional regulatory networks}
\label{sec:causal-gene-networks}

\begin{figure}[t!]
\begin{center}
\includegraphics[width=.95\textwidth]{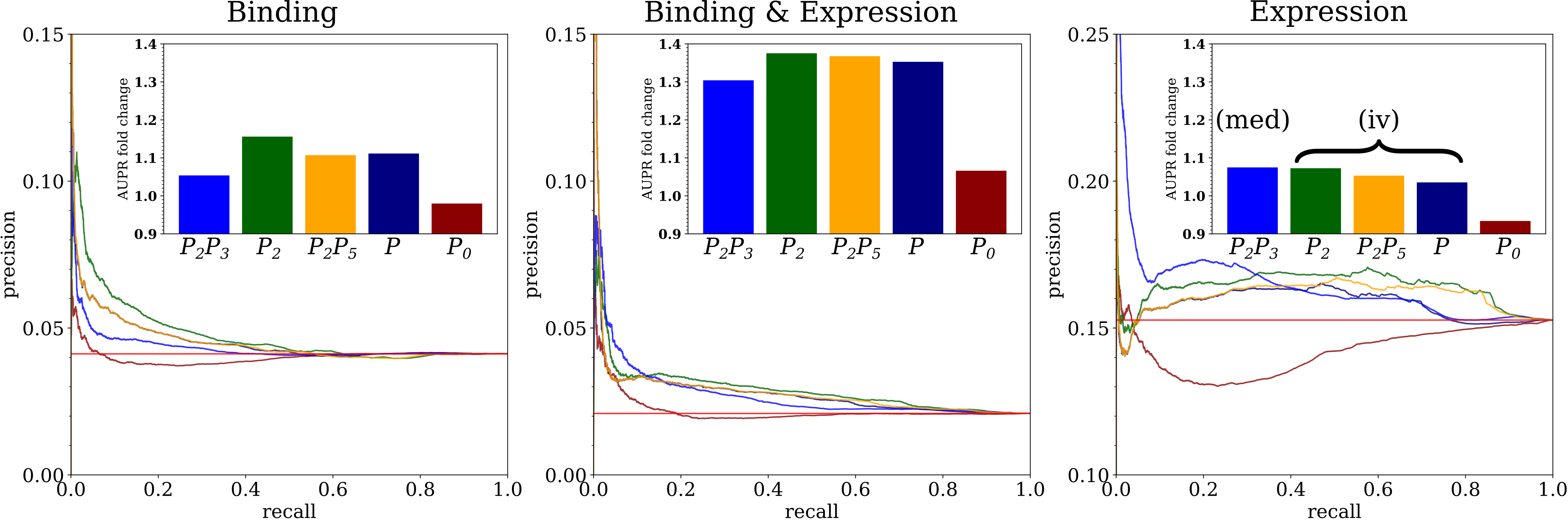}
\end{center}
\caption{\textbf{Performance of causal inference on \textsc{Yeastract} ground truths.} 
    Precision-recall curves for four causal inference methods ($P_2P_3$, $P_2$, $P_2P_5$, $P$) and one coexpression method ($P_0$) are shown  for the  Binding (left), Binding and Expression (center) and Expression (right) ground-truth networks. The insets show the area under the precision-recall curves (AUPR) as the the ``fold change'' relative to the baseline performance for random predictions.
    \ced{The horizontal red line shows the baseline performance for random predictions and is used as reference for AUPR fold change (insets).}
    The network inference methods are described in Section~\ref{sec:findr-reconstr}.
    }
\label{fig:PR}
\end{figure}

We assessed the performance of networks predicted by Findr on three ground-truth networks of transcriptional regulatory interactions in yeast, where targets of a transcription factor (TF) are defined by TF-DNA binding interactions (``Binding'' network), differential expression upon TF perturbation (``Expression'' network), or the intersection of them (``Binding \& Expression'' network) (see Methods and Table~\ref{tab:ground_truth}). The precision-recall curves for the four causal inference methods showed the characteristic peak of high precision at low recall indicative of an enrichment of true positives among the predictions with highest posterior probabilities, and confirmed by increased area under the precision-recall curve (AUPR) compared to random predictions (Figure~\ref{fig:PR}). This was markedly the case for the Binding \& Expression ground-truth, with AUPR more than 1.3 times higher than random. This is consistent with the notion that genes that are bound by a TF as well as differentially expressed upon TF perturbation are more likely to be real TF targets, that is, that the Binding \& Expression ground-truth is of higher quality than the others. Differences between causal inference methods were modest, with instrumental variable methods ($P_2$, $P_2P_5$, $P$) showing somewhat better performance than the mediation-based method ($P_2P_3$) on the Binding and Binding \& Expression ground-truths, and vice versa on the Expression ground-truth (Figure~\ref{fig:PR}).
In contrast to the causal inference methods, the coexpression-based method ($P_0$) did not show any improvement over random predictions. This is not surprising. An unbiased evaluation of 35 diverse methods for network inference from expression alone did not find any improvement over random predictions on a comparable ground-truth network for yeast \cite{marbach2012predictive}.

\subsection{The performance of mediation \ced{saturates} at large sample size}
\label{sec:medi-large-sample}

\begin{figure}[th]
\begin{center}
\includegraphics[width=.95\textwidth]{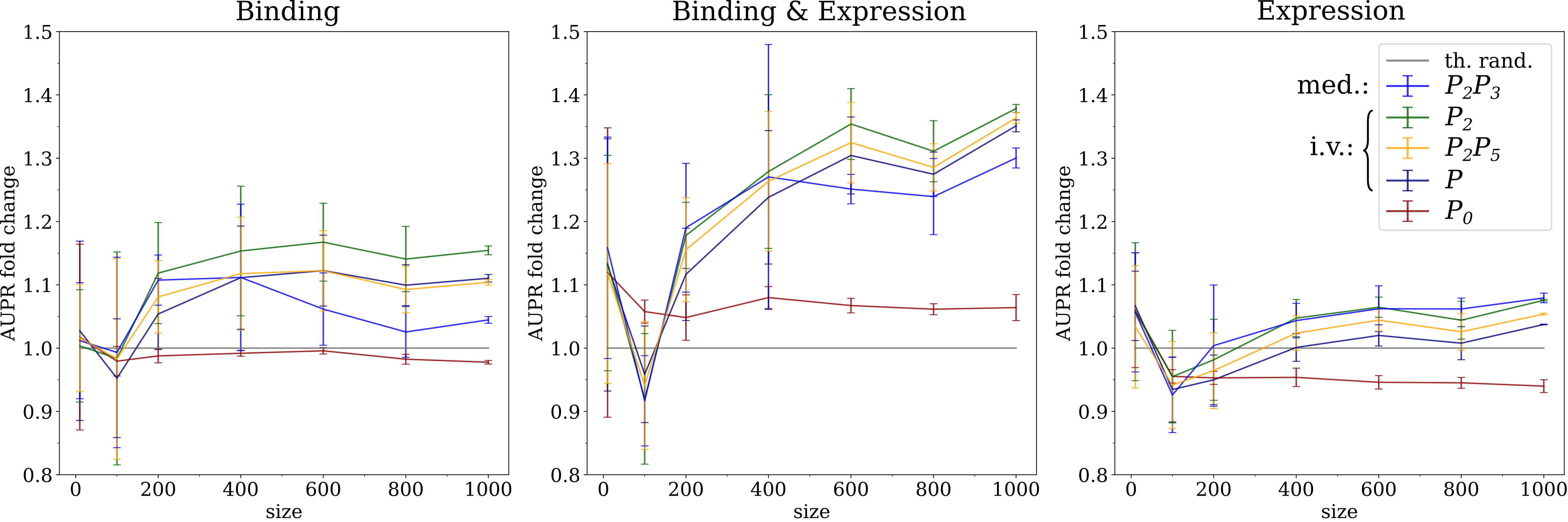}
\end{center}
\caption{\textbf{Performance of causal inference across sample sizes.} 
    AUPR fold change values for four causal inference methods ($P_2P_3$, $P_2$, $P_2P_5$, $P$) and one coexpression method ($P_0$) (see Section~\ref{sec:findr-reconstr}) at various sample sizes for the  Binding (left), Binding \& Expression (center) and Expression (right) ground-truth networks. Four samples were randomly drawn from the expression data and evaluated with each test. Error bars represent the standard deviation across the four subsets. The horizontal grey line indicates the level of random predictions. The fold change is relative to the  baseline performance for random predictions.
    }
\label{fig:subsampling}
\end{figure}

The availability of more than 1,000 segregants in the genotype and gene expression dataset allowed us to evaluate the performance of network inference across sample sizes by random subsampling of the data. The clearest pattern was again observed for the Binding \& Expression ground-truth, consisting of the most reliable known transcriptional regulatory interactions, where the three instrumental variable methods ($P_2$, $P_2P_5$, $P$) showed a monotonous increase in AUPR with increasing sample size (Fig.~\ref{fig:subsampling}). The mediation based method ($P_2P_3$) initially showed a similar performance as the instrumental variable methods, \ced{but saturated above 400 samples when accounting for statistical error and dropped from having the highest to the lowest average performance of all causal inference methods.} \cedst{reached its best AUPR around 600 samples and declined after that.} The same \cedst{phenomena of increasing performance of instrumental variable methods, and decreasing performance of the mediation-based method at large sample sizes} \ced{pattern} is also observed on the Binding ground-truth, albeit in a less pronounced way, presumably due to lower AUPR values relative to random predictions for all methods.

\ced{These results are consistent with previous work on simulated data, where we observed a decrease in} \cedst{The phenomenon of decreasing} performance with increasing sample size for mediation-based methods \cedst{was also observed in simulated data} \cite{wang2017}. \cedst{w}\ced{T}here we showed that hidden confounders and measurement noise can lead to a residual correlation between the causal anchor $E$ and a target gene $B$ after adjusting for the regulatory gene $A$ (cf.\ Fig.~\ref{fig:tests-models}). At sufficiently large sample size, this residual correlation becomes significantly different from zero  and thereby leads to a false negative prediction.

Sample size showed little effect on the coexpression method $P_0$ for sample sizes larger than 400 for all ground truths. This is consistent with the notion that estimates of correlations will stabilize around their true values at smaller sample sizes than estimates of causal effects.

\subsection{Instrumental variable methods are affected by genomic linkage blocks}
\label{sec:iv-genomic-linkage}

Next, we assessed the extent to which instrumental variable methods are affected by genomic linkage between causal anchors which would lead to false positive predictions due to (real or apparent) pleiotropic effects (cf.\ Fig.~\ref{fig:tests-models}). For instance for the $P_2$ method, if two genes in the same genomic neighbourhood have causal anchors with strongly correlated genotype values, the method would predict them to have similar sets of target genes.

To perform the analysis, we first truncated the posterior probability values in order to obtain discrete, directed networks. Thresholds were determined to obtain networks with an expected FDR $\leq 5\%$ (for the instrumental variable methods) or $\leq 10\%$ (for the mediation-based method) (see Methods section). The larger FDR value for mediation was chosen to counterbalance its increased false negative rate, and resulted in posterior probability thresholds that were comparable between all methods (Table \ref{tab:network_properties}).

\begin{table}[th]
\begin{center}
\begin{tabular}{l lc rrr r}
\\\hline  Test      & $p_\mathrm{th}$ & FDR  &  $N_R$    & $N_T$     & $N_E$       & $\rho$   %
\\\hline
        $P_2P_3$    &  0.8175	&	0.09953 &   1,808   &  5,628    &     144,091 & 0.014  %
\\	    $P_2$	    &  0.825	&	0.04974 &   2,884   &  5,720	&   2,319,854 & 0.141  %
\\	    $P_2P_5$	&  0.8375	&	0.04994 &   2,884   &  5,719    &   1,740,251 & 0.106  %
\\	    $P$	        &  0.8575	&	0.04982 &   2,884   &  5,720	&   2,428,039 & 0.147  %
\\\hline
\end{tabular}
\end{center}
\caption{\textbf{Properties of thresholded predicted networks.}
    We report the thresholds ($p_\mathrm{th}$) used to select significant interactions 
    for the four causal inference methods, the corresponding global False Discovery Rate (FDR),
    as well as 
    descriptors for the resulting networks:
    $N_R$ is the number of regulating genes, 
    $N_T$ is the number of target genes, 
    $N_E$ is the number of edges,
    and $\rho$ is the filling ratio of the adjacency matrix,
    \ced{i.e.\ the ratio of non-zero and zero values in the thresholded matrices.}
    }
\label{tab:network_properties}
\end{table}

\begin{figure}[th]
\begin{center}
\includegraphics[width=.98\textwidth]{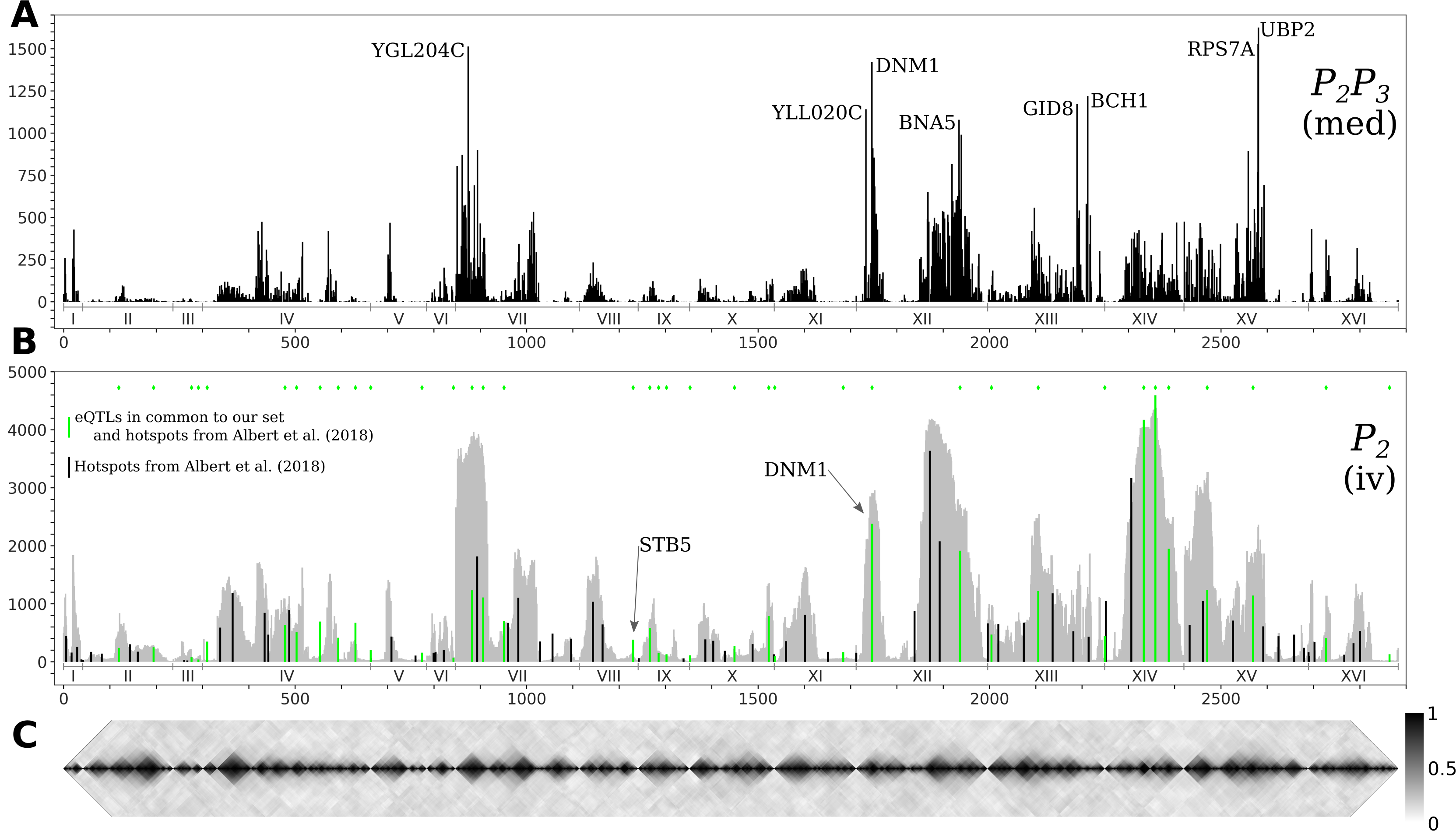}
\end{center}
\caption{\textbf{Hotspots and genotype covariance.}
    \textbf{A.} The counts of significant interactions for the mediation-based method $P_2P_3$ at FDR below 10\%, with annotations for eight regulatory genes with more than 1,000 targets. \textbf{B.} The counts of significant interactions for the instrumental variable method $P_2$ at FDR below 5\% (in grey), and the number of non-zero effects for 102 hotspot markers from \cite{albert2018genetics} (in black); the subset of these hotspots that are also a causal anchor (i.e.\ the strongest local eQTL for at least one gene) for the \findr{} analysis are marked in green and are also indicated by diamonds at the top of the panel. Interaction count plots for the other instrumental variable methods are in Supp.\ Fig.~\ref{fig:hotspots_sup}. \textbf{C.}  The diagonal of the genotype covariance matrix for the 2884 causal anchor eQTLs.
    Genes are ordered along the horizontal axis according to the position of their causal anchor in the yeast genome.
    }
\label{fig:hotspots}
\end{figure}

Despite the similar posterior probability thresholds, the instrumental variable networks are around ten times more densely connected than those obtained by the mediation-based method (Table~\ref{tab:network_properties}); a difference that cannot be explained by the lower sensitivity of the latter alone. 
We show that in the instrumental variable networks, high interaction counts occur in blocks that roughly follow  the structure of the causal-anchor genotype covariance, whereas they occur more in spikes in the mediation network (Fig.~\ref{fig:hotspots}). This becomes apparent when plotting the number of targets for each regulatory gene (i.e.\ each gene with a significant \emph{cis}-eQTL) \textit{versus} its position on the genome.
In instrumental variable methods, the genomic causal anchor is used as a ``proxy'' for the regulatory gene. Hence, if the causal anchor genotypes of two genes within the same locus are correlated due to genomic linkage, then their target sets will unavoidably be similar as well, resulting in the pattern observed in Fig.~\ref{fig:hotspots}. In mediation-based methods, the expression profile of the regulatory gene is used as the mediator (in test 3, cf.\ Fig.~\ref{fig:tests-models}A), and hence a target set will be specific to a regulator, even when its causal anchor is correlated or shared with other genes.

\subsection{Causal network inference suggests causal genes for transcriptional hotspots}
\label{sec:caus-netw-infer}

Regions of the genome that are statistically associated with variation in expression of a high number of genes (the peaks in Fig.~\ref{fig:hotspots}B) are called transcriptional ``hotspots'', and finding the causal genes underlying a hotspot is an important problem in quantitative genetics \cite{rockman2006genetics}. Albert \textit{et al.} \cite{albert2018genetics} identified 102 hotspot loci using their data, and developed a fine-mapping strategy to narrow the confidence intervals for the hotspot locations. Overlaying the hotspot markers (median bootstrap hotspot locations \cite{albert2018genetics}) with the $P_2$ target counts in the 5\% FDR network (Fig.~\ref{fig:hotspots}B) showed good consistency, as expected; 37 of those hotspot markers were in our list of causal anchors (i.e.\ strongest local eQTL for at least one gene). Albert \textit{et al.} \cite{albert2018genetics} defined candidate causal hotspot genes as the genes located within the fine-mapped hotspot regions, and for 26 hotspots they obtained three or fewer candidate genes. Here we illustrate how causal gene network inference can contribute to the identification of causal hotspot genes using two representative examples \textit{STB5} and \textit{DNM1}.

\begin{figure}[th]
\begin{center}
\includegraphics[width=.95\textwidth]{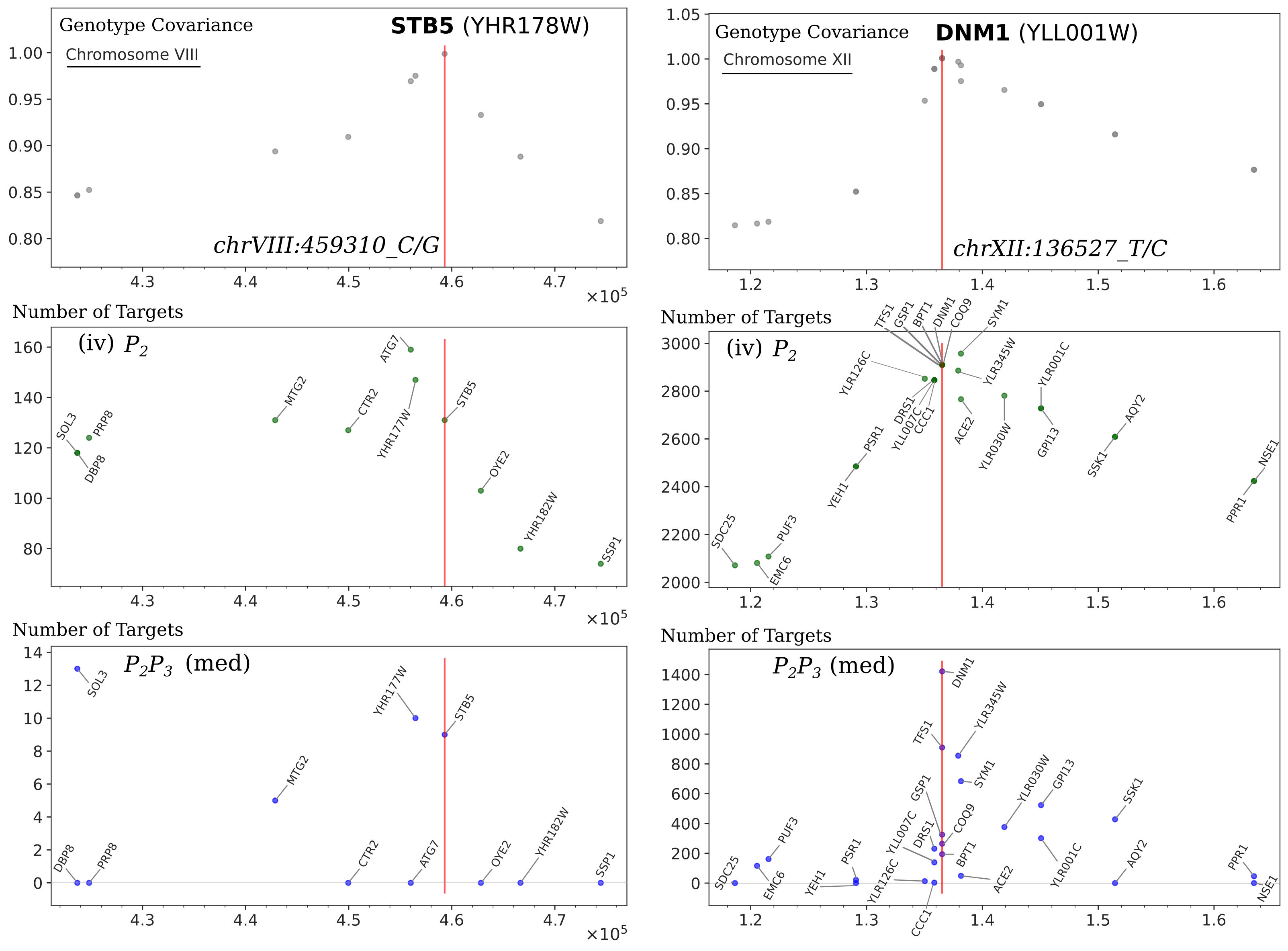}
\end{center}
\caption{\textbf{Details of predicted targets in the vicinity of two genes.}
    We show the local structure at two genes:
     \textit{STB5} with eQTL {chrVIII:459310\_C/G} \textbf{(left)}
      and 
     \textit{DNM1} with eQTL {chrXII:136527\_T/C} \textbf{(right)}.
    The \textbf{top row} shows genotype covariance in the vicinity of the eQTL (red line) for the gene,
       in the region where the covariance is greater than $0.8$.
    The \textbf{middle row} shows number of targets predicted by $P_2$ \ced{(instrumental variables)} at FDR below 5\%.
    The \textbf{bottom row} shows number of targets predicted by $P_2P_3$ \ced{(mediation)} at FDR below 10\%.
    The horizontal axis gives the position along the chromsome of the eQTL corresponding to each gene.
    Genes are annotated with their short name where available.
    Note that data points overlap in genotype covariance and in $P_2$
        for some genes because they share the same eQTL
         and that $P_2P_3$ gives no targets on certain genes.
    }
\label{fig:hotspots_zooms}
\end{figure}

\textit{STB5} is a transcription activator of multidrug resistance genes \cite{kasten1997identification}, and the only gene located in one hotspot region on chromosome VIII. The hotspot marker, chrVIII:459310\_C/G, is located 11 bp upstream from \textit{STB5}, and is the causal anchor for \textit{STB5} and for no other genes (Fig.~\ref{fig:hotspots}B and Fig.~\ref{fig:hotspots_zooms}~Left). The instrumental variable method $P_2$ predicted 131 targets at FDR below 5\% for \textit{STB5}, which are strongly enriched for \textit{STB5} targets in the Binding (hypergeometric $p$-value $2.3\cdot 10^{-12}$) and Binding \& Expression (hypergeometric $p$-value $1.9\cdot 10^{-10}$) ground-truth networks. This suggests that when a hotpspot can be confidently mapped to a single gene, instrumental variable methods predict biologically plausible target sets confirming the candidate causal hotspot gene. In contrast, the mediation-based method $P_2P_3$ predicted only nine \textit{STB5} targets at FDR below 10\%, with no enrichment in the ground-truth networks. A possible mechanism that could explain the loss of sensitivity of mediation in this case was already suggested by Albert \textit{et al.}\  \cite{albert2018genetics}: \textit{STB5} does not show allele-specific expression, but carries protein-altering variants between the two yeast strains that were crossed, suggesting that the causal variants in this hotspot act by directly altering Stb5p protein activity; moreover Stb5p is predicted to target its own promoter in \textsc{Yeastract}. Taken together, this leads to a model where transcription of \textit{STB5} is a noisy measurement of Stb5p level, that does not block the path from the protein-altering variants to \textit{STB5} target genes via Stb5p protein level (Supp.\ Fig.~\ref{fig:stb5_sup}). Hence conditioning on \textit{STB5} transcription in $P_2P_3$ does not remove the association between these variants and the target genes completely, resulting in false negative predictions by a process similar to the measurement noise model studied in \cite{wang2017}.

\textit{DNM1} is a gene located near a hotspot region on chromosome XII, and is among the genes with highest target count in the mediation-based network   (Fig.~\ref{fig:hotspots}A, Fig.~\ref{fig:hotspots_zooms}~Right). The hotspot marker is also the causal anchor of \textit{DNM1}, which is located 11,363 bp downstream of this marker and outside the hotspot region mapped by Albert \textit{et al.} Comparison with the target counts in the instrumental variable network, which closely follow the genotype covariance pattern, shows that \textit{DNM1} is the gene in this region that retains the most targets by far in the mediation network ($P_2$, 2910 targets; $P_2P_3$, 1421 targets; Fig.~\ref{fig:hotspots_zooms}~Right). This is particularly true when compared with two of the four candidate causal genes of Albert \textit{et al.}\ that also have a local eQTL within the hotspot region, \textit{YLL007C} (also known as \emph{LMO1}) ($P_2$, 2846 targets, $P_2P_3$, 139 targets) and \textit{MMM1}
($P_2$, 3610 targets; $P_2P_3$, 8 targets).
  Based on the high specificity of the $P_2P_3$ test, we conjecture that \textit{DNM1} is a more likely causal gene for this hotspot than \textit{LMO1} or \textit{MMM1}. Functional analysis in this case does not help to distinguish between these candidates, because Dnm1p and Mmm1p are both essential proteins for the maintenance of mitochondrial morphology \cite{otsuga1998dynamin}, and Lmo1p is a signaling protein involved in mitophagy \cite{schmitz2015identification}. However, deletion of \emph{DNM1} and \emph{MMM1} results in distinct mitochondrial phenotypes \cite{otsuga1998dynamin}, and hence this prediction is experimentally testable in principle.

\section{Discussion}
\label{sec:discussion}

\subsection{Causal inference from genomics and transcriptomics data infers truly directed gene networks}
\label{sec:caus-infer-from}

Reconstructing transcriptional regulatory networks from transcriptomics data has been a major research focus in computational biology during the last 20 years. Existing methods span the entire range of correlation, mutual information, regression, Bayesian networks, random forest, and other machine learning methods, as well as meta-methods combining multiple of these aproaches \cite{marbach2012wisdom}. Yet, performance on eukaryotic gene expression data has been disappointing, with overlap between predicted and known networks generally not better than random predictions \cite{marbach2012wisdom}. To some extent, this is due to the lack of reliable ground-truth data. For instance, there is little overlap between the two most common high-throughput experimental techniques for measuring regulatory interactions, mapping TF-DNA binding sites using ChIP-sequencing and measuring differential expression after TF deletion or overexpression \cite{cusanovich2014functional}, see also Table~\ref{tab:ground_truth}.  Exceptions to this rule are the transcriptional networks controlling early development in multi-cellular organisms, which are mapped in exquisite detail in some model organisms \cite{macarthur2009developmental}. When conventional network inference methods are applied to developmental transcriptome data, good performance is in fact observed \cite{joshi2015}. Nevertheless the problem of reconstructing \emph{signalling} transcriptional networks from observational expression data remains, and a key missing ingredient in existing approaches is the \emph{directionality} of the edges. Without additional information, any association inferred from transcriptomics data alone is essentially symmetric.

Causal inference is designed to reconstruct truly directed networks, by integrating genomics and transcriptomics data based on general principles of quantitative trait locus analysis \cite{jansen2001genetical}. The publication of a dataset of more than 1,000 yeast segregants has allowed us to  demonstrate that causal inference indeed results in directed networks with strong, non-random overlap with networks of known transcriptional interactions. Moreover, the overlap was highest for the most reliable ground-truth that combined two sources of experimental evidence (DNA binding and response to perturbation). Although 1,012 samples for an organism with around 6,000 genes may seem a large number, our analysis also shows that there is no sign yet that performance is saturating as a function of sample size. Causal inference indeed requires larger sample sizes than coexpression-based methods, because it relies on more subtle patterns in the data, something that was already apparent in early considerations of causal inference in this context \cite{li2010critical}.

Although the integration of genomics and transcriptomics data addresses the key shortcoming pertaining to lack of directionality in network inference when using transcriptomics data alone, important limitations remain. Apart from those already discussed at length in this paper---low sensitivity due to hidden confounders for mediation-based methods, and increased false positive rate due to genomic linkage for instrumental variable methods---another fundamental problem remains: transcriptional regulation is, for the most part, carried out by proteins. \ced{Hence, causal interactions inferred from genomics and transcriptomics data are by definition indirect. If an intermediate, unmeasured protein $C$ (e.g., the protein product of $A$) lies on the path from gene $A$ to gene $B$, that is, $A$ also mediates associations between $C$ and local eQTLs for $A$, then this does not affect the causal inference for the interaction $A\to B$ (Supp.\ Fig.~\ref{fig:models_sup}). However,} variation in transcription level of a transcription factor (or other regulatory protein) does not always translate to equal variation in protein level, and \textit{vice versa}.
For instance, Albert \textit{et al.}\ \cite{albert2018genetics} found several protein-altering variants in candidate causal genes mapped to hotspot regions that did not have any local eQTLs.  In such cases, our methods would wrongly assign the \textit{trans}-associated target genes to a gene with local eQTL (if one exists), and  miss the non-varying (at transcription level) causal gene. This limitation can only be addressed by integrating another layer of information---proteomics data, which  is not yet available in comparable sample sizes.

\ced{The methods implemented in \findr\ and analyzed in this paper are broadly representative of the current state-of-the-art for causal inference from genomics and transcriptomics data. Nevertheless, some ideas have been proposed recently that we did not evaluate here. For instance, in addition to the statistical tests implemented in \findr, Badsha and Fu \cite{badsha2019learning} propose to also use a causal anchor for the target gene $B$ to obtain evidence for a causal interaction $A\to B$. However, including this test requires limiting the analysis to interactions where both source and target genes have a significant eQTL. Yang \emph{et al.}\cite{yang2017identifying} on the other hand propose to address the hidden confounder problem in mediation by adjusting for selected surrogate variables (e.g.\ principal components). However, such variables are necessarily composed of combinations of genes and it is challenging to ensure that they only represent common parents and no common children of an $A\to B$ interaction (which would introduce false positives if conditioned upon, see Supp.~Fig.~\ref{fig:models_sup}). It will be of interest to include these developments in future comparisons.}

\subsection{Biological data matches theoretical predictions}
\label{sec:exper-data-match}

Causal inference is in essence a hypothesis-driven approach: the causal diagrams in Figure~\ref{fig:tests-models} encode prior knowledge and assumptions of how genotypes, genes, and unknown confounding factors influence each other. Based on these diagrams, we can make certain predictions about the patterns we expect to find in the data, such as the relative sensitivity and specificity of mediation \textit{versus} instrumental variable methods, the different situations where each method will be successful or not, etc. It is gratifying to see these predictions confirmed using real data, strengthening significantly our previous findings on simulated data \cite{wang2017}.

The hypothesis-driven nature of causal inference lies in between the use of biophysical models of gene regulation and the application of \cedst{``black box''} \ced{unsupervised} machine learning methods for reconstructing gene regulatory networks. Biophysical approaches attempt to include quantitative models of TF-DNA interactions into the network inference process  \cite{bussemaker2007,aijo2016biophysically}, but are hampered by a lack of resolution in omics data (due to both noise within a sample, and limited sampling density). \cedst{``Black box''} \ced{Unsupervised} machine learning approaches search for non-random patterns within the data, but \ced{without prior specification} of the type of pattern being sought, and they lack the \cedst{focus} \ced{ability} to identify truly directed associations. \ced{Supervised or semi-supervised methods could potentially overcome this limitation, but require data of known interactions for training, which is sparse for non-model organisms \cite{maetschke2014supervised}.}

The agreement between theoretical predictions \ced{and empirical results} indicates that we have a correct understanding of how causal gene network inference algorithms work, and how to interpret results in terms of what type of interactions these algorithms do and do not identify, albeit without any reference to the underlying biophysical mechanisms.

\subsection{Practical recommendations \ced{and future work}}
\label{sec:pract-recomm}

We conclude this paper by sharing practical recommendations for researchers wanting to apply causal inference methods for the integration of genomics and transcriptomics data.

In general, we recommend instrumental variable over mediation-based methods, as their increased sensitivity tends to outweigh the higher specificity of mediation-based methods. The \cedst{decrease} \ced{saturation} of performance of mediation-based methods with increasing sample sizes is particularly worrying, although for most current datasets the point where performance \cedst{begins to decrease} \ced{saturates} is probably not yet reached.

We found limited differences between instrumental variable methods. In the absence of any ground-truth data to evaluate results, we would generally recommend to use the $P_2P_5$ method, because it will remove at least the most obvious cases of pleiotropy from $P_2$, while having an easier interpretation than the $P$ method.

The main weakness of instrumental variable methods is their susceptibility to false positive predictions due to genomic linkage. This is a particular concern in data from experimental crosses or inbred organisms, where linkage blocks are large. However, also in human data it has been found that around 10\% of non-redundant local eQTLs are associated to the expression of multiple nearby genes \cite{tong2017shared}. As illustrated, mediation-based causal inference and manual analysis of gene function can sometimes be used to resolve linkage of causal anchors.

In conclusion, causal inference from genomics and transcriptomics data is a more powerful approach for reconstructing causal gene networks than using transcriptomics data alone, which could be further improved by \ced{the inclusion of additional layers of omics data and} the development of methods \ced{to control for or find signal in residual correlations among genes in mediation analyses}, and
to resolve genomic linkage and pleiotropic effects from transcriptional hotspots \ced{in instrumental variable analyses}.

\bibliographystyle{erc_long} 
\bibliography{bibsysbiol,references_yeast}


\newpage


\renewcommand\thesection{S\arabic{section}}
\renewcommand\thefigure{S\arabic{figure}}
\renewcommand\thetable{S\arabic{table}}
\renewcommand\theequation{S\arabic{equation}}
\setcounter{figure}{0}
 \setcounter{table}{0}
 \setcounter{section}{0}
\setcounter{equation}{0}

\begin{center}
  {\LARGE \textbf{Supplementary Information}}
\end{center}

\begin{table}[bth]
\begin{center}
\begin{tabular}{lll}
\\\hline  Method & $p_\mathrm{th}$ & FDR
\\\hline
    $P_2P_3$	&	0.8175	&	0.09953
\\  $P_2$	    &	0.825	&	0.04974
\\  $P_2P_5$	&	0.8375	&	0.04994
\\	$P$	        &	0.8575	&	0.04982
\\	$P_0$	    &	0.86	&	0.00986
\\\hline
\end{tabular}
\end{center}
\caption{\textbf{FDR thresholds.}
    The thresholds ($p_\mathrm{th}$) reported here  were used to select significant interactions
    for the methods  shown in figure \ref{fig:hotspots} and \ref{fig:hotspots_sup}.
    }
\label{tab:fdr_sup}
\end{table}

\begin{figure}[th]
\begin{center}
\includegraphics[width=.95\textwidth]{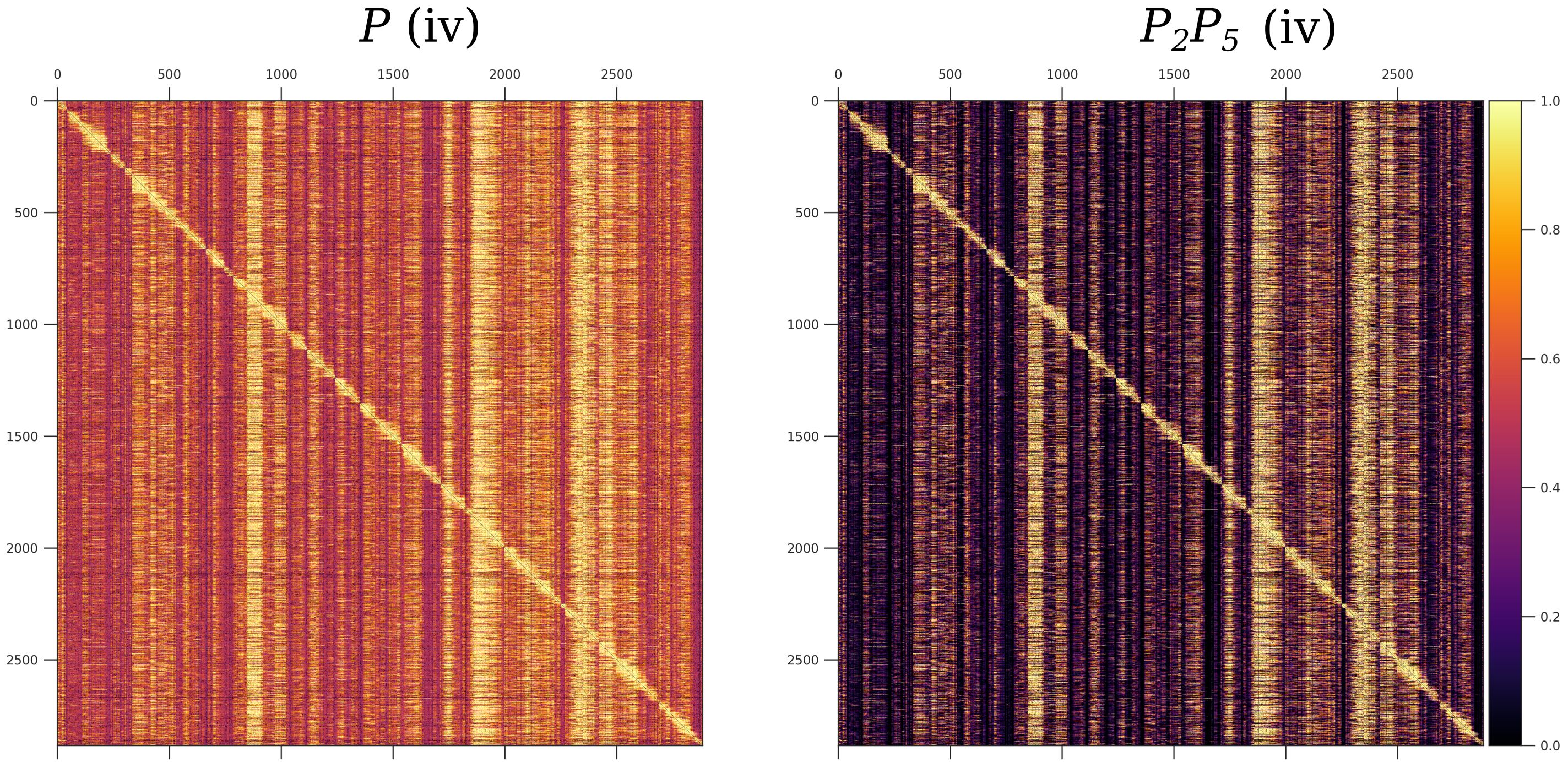}
\end{center}
\caption{\textbf{Matrices of predicted gene interactions.}
    These square matrices represent the interactions between 2884 genes with causal anchors (eQTLs),
        posterior probability values are color coded.
        Vertical bands correspond to hotspots.
    \textbf{Left}: The instrumental variable test with partial pleiotropy $P$.
    \textbf{Right}: The instrumental variable test with perfect pleiotropy $P_2P_5$.
    The genes are ordered according to the position of their causal anchor in the full yeast genome.
    Definitions of the tests are given in the Methods section.
    This figure complements Fig. \ref{fig:matrices} in the main text.
    }
\label{fig:matrices_sup}
\end{figure}

\begin{figure}[th]
\begin{center}
\includegraphics[width=.5\textwidth]{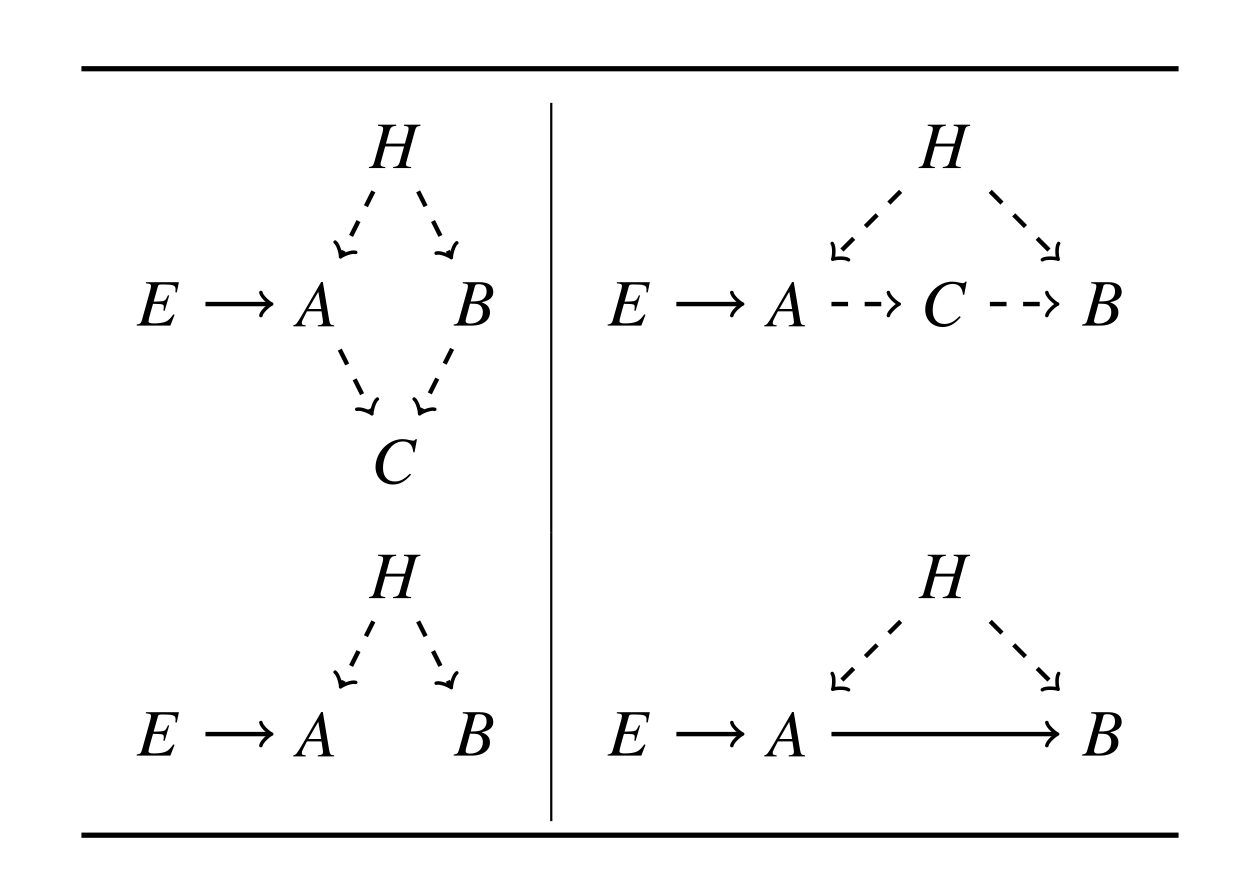}
\end{center}
\caption{\textbf{Comparison of causal models.}
    These graphical models illustrate possible interactions between genes.
    The hidden confounder $H$ would be a common parent of $A$ and $B$ as shown also in Fig. \ref{fig:tests-models}[B].
    The upper row shows a possible ground truth, the lower row shows how this scenario would be classified by \findr.
    \textbf{Left}: A common child $C$ of $A$ and $B$ does not affect the predictions of \findr\, regardless of the presence of a common parent $H$.
    \textbf{Right}: An intermediary node $C$ could mediate the interaction between $A$ and $B$, regardless of a common parent $H$.
    }
\label{fig:models_sup}
\end{figure}

\begin{figure}[th]
\begin{center}
\includegraphics[width=.95\textwidth]{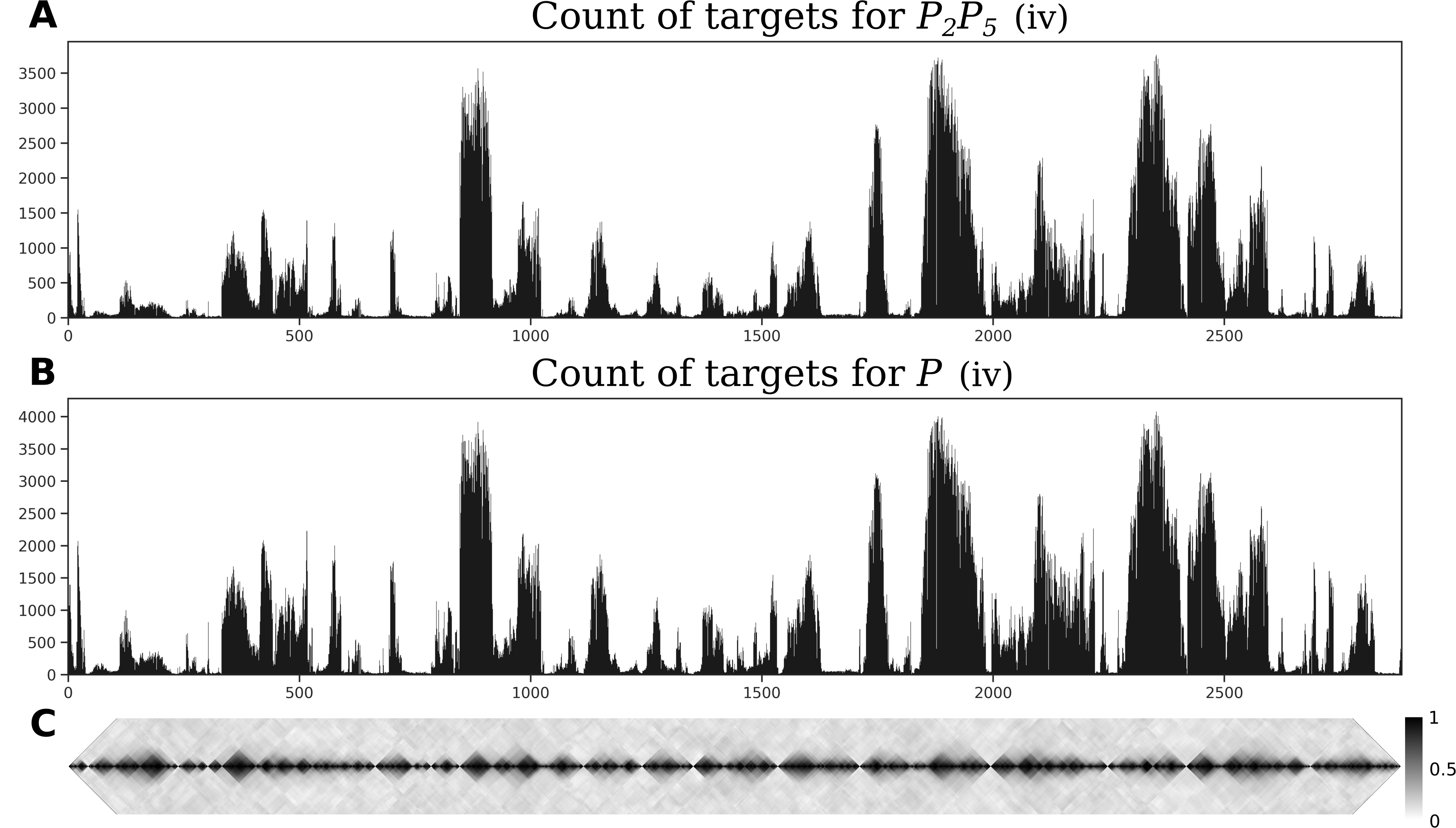}
\end{center}
\caption{\textbf{Hotspots and genotype covariance.}
     \textbf{A} and \textbf{B} show the counts of significant interactions
     for two inference methods.
    Genes are ordered along the horizontal axis according to the position of their causal anchor in the full yeast genome.
    \textbf{A}: instrumental variables with perfect pleiotropy ($P_2P_5$) at FDR 5\%.
    \textbf{B}: instrumental variables with partial pleiotropy ($P$) at FDR 5\%.
    The thresholds used are reported in Tab. \ref{tab:fdr_sup}.
    \textbf{C}:  The diagonal of the genotype covariance matrix for the 2884 eQTLs.
    }
\label{fig:hotspots_sup}
\end{figure}

\begin{figure}[th]
  \centering
  \includegraphics[width=.5\textwidth]{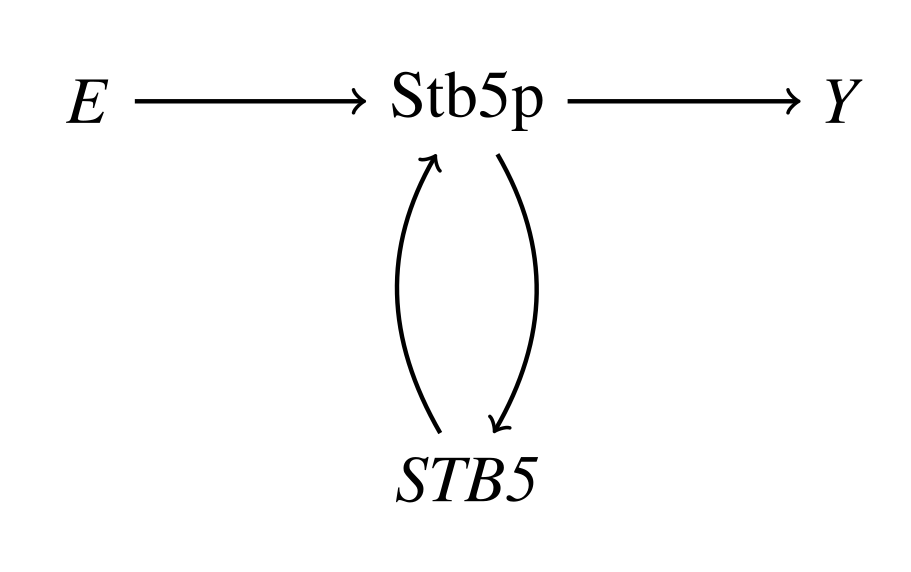}
  \caption{\textbf{Hypothetical model for the \textit{STB5} hotspot.} Stb5p protein level is determined by \textit{STB5} transcription level and the genotype of one or more protein-altering variants $E$, and in turn affects \textit{STB5} transcription level by an auto-regulatory loop. Expression of \textit{STB5} target genes $Y$ is determined by \textit{STB5} transcription only through Stb5p level. Even in the absence of any hidden confounders, \textit{STB5} transcription does not block the path between $E$ and $Y$, and unless the correlation between \textit{STB5} transcription and Stb5p level is perfect (no biological or experiment noise), conditioning on \textit{STB5} transcription level will not remove the statistical association between $E$ and $Y$. This model is consistent with the observed lack of allele-specific expression of \textit{STB5} \cite{albert2018genetics}, and with the fact that the instrumental variable method $P_2$ correctly identifies target genes with Stb5p binding sites, but the mediation-based method $P_2P_3$ does not.}
  \label{fig:stb5_sup}
\end{figure}

\end{document}